\documentclass[aps,prx,twocolumn,amsmath,amssymb,floatfix,superscriptaddress,longbibliography]{revtex4-1}
\usepackage{natbib}

\usepackage{graphicx}
\usepackage{dcolumn}
\usepackage{bm}
\usepackage{mathrsfs}
\usepackage{amsmath}
\usepackage[linesnumbered, ruled, vlined]{algorithm2e}
\usepackage{float}
\usepackage{subfigure}
\usepackage{enumitem}
\usepackage{color}
\usepackage{cancel}

\newcommand{\abs}[1]{\left.|#1|\right.}
\newcommand{\tabincell}[2]{\begin{tabular}{@{}#1@{}}#2\end{tabular}}

\begin{document}
	
	\title{Circuit-Depth Reduction of Unitary-Coupled-Cluster Ansatz by Energy Sorting}
	
	
	\author{Yi Fan}
	\affiliation{Central Research Institute, 2012 Labs, Huawei Technologies}
	\affiliation{Hefei National Laboratory for Physical Sciences at the Microscale, University of Science and Technology of China, Hefei, Anhui 230026, China}
	\author{Changsu Cao}
    \affiliation{Tsinghua University, Department of Chemistry, Beijing 100084, China}

	\author{Xusheng Xu}
    \affiliation{Central Research Institute, 2012 Labs, Huawei Technologies}
	\author{Zhenyu Li}
	\affiliation{Hefei National Laboratory for Physical Sciences at the Microscale, University of Science and Technology of China, Hefei, Anhui 230026, China}
	
	\author{Dingshun Lv} \email{ywlds@163.com}
	\affiliation{Central Research Institute, 2012 Labs, Huawei Technologies}
	\author{Man-Hong Yung} \email{yung.manhong@huawei.com}
	\affiliation{Central Research Institute, 2012 Labs, Huawei Technologies}

	\date{\today}
	
	\begin{abstract}
		Quantum computation represents a revolutionary approach for solving problems in quantum chemistry. However, due to the limited quantum resources in the current noisy intermediate-scale quantum (NISQ) devices, quantum algorithms for large chemical systems remains a major task. In this work, we demonstrate that the circuit depth of the unitary coupled cluster (UCC) and UCC-based ansatzes in the algorithm of variational quantum eigensolver can be significantly reduced by an energy-sorting strategy. Specifically, subsets of excitation operators are first pre-screened from the operator pool according to its contribution to the total energy. The quantum circuit ansatz is then iteratively constructed until the convergence of the final energy to a typical accuracy. For demonstration, this method has been successfully applied to molecular and periodic systems. Particularly, a reduction of 50\%$\sim$98\% in the number of operators is observed while retaining the accuracy of the origin UCCSD operator pools. This method can be straightforwardly extended to general parametric variational ansatzes.
	\end{abstract}

	\maketitle
	
	\section{\label{intro}Introduction}
	
	Noisy intermediate-scale quantum(NISQ) devices have been studied extensively to demonstrate a wide variety of small-scale quantum computations\cite{preskill2018quantum}. The ground-state problem of quantum chemistry is one of the applications that can potentially achieve demonstrable quantum advantage on quantum computers \cite{feynman_simulating_1982,abrams_quantum_1999}. 
    Quantum phase estimation (QPE) and variational quantum eigensolver (VQE) are two dominant algorithms to solve for the ground-state of a chemical system.
	QPE implements the time evolution operator $ U = e^{i \hat{H} t}$ and evolves the Hamiltonian of the molecule in time on a quantum computer \cite{aspuru2005simulated} to obtain the energy spectrum with accuracy comparable to full configuration interaction (FCI)\cite{cao2018quantum, expri-nmr-1, expri-nmr-2, lanyon2010towards, expri-qp-1, expri-nv-1, omalley_scalable_2016}. However, accurate QPE simulation requires long coherence time, high (two-qubit) gate fidelity and even error correction devices, which is far beyond the NISQ era.
	In contrast to QPE, VQE\cite{Yung2014, peruzzo_variational_2014} embeds quantum simulations into a classical optimization process, which yields a much shallower circuit and relieves the requirement of coherence time and gate fidelity. Therefore, VQE is thus preferable for NISQ devices and has been experimentally demonstrated on leading quantum platforms including photonic quantum processors\cite{peruzzo_variational_2014}, superconducting devices\cite{omalley_scalable_2016, kandala_hardware-efficient_2017,colless_computation_2018,expri-symmetry-1} and trapped-ion quantum processors\cite{expri-vqe-1,hempel_quantum_2018,nam_ground-state_2019}.
 
	In the procedure of VQE, a trial wave function encoded into a parametric quantum circuit. On the quantum computer, the energy is estimated by measuring the expectation values of the electronic Hamiltonian; on the classical computer, the parameters are optimized and updated to minimize the energy. 
    The main ingredient of VQE is the wave function ansatz. Common schemes include physically motivated ansatz (PMA) based on systematic techniques to approximate the exact wave function, such as the unitary coupled-cluster theory or UCC-based ansatzes\cite{omalley_scalable_2016, expri-vqe-1,hempel_quantum_2018,nam_ground-state_2019,ryabinkin2020iterative, kawashima2021efficient}, and hardware heuristic ansatz (HHA) which considers specific hardware structure and employs entangling blocks, such as the hardware efficient ansatz\cite{kandala_hardware-efficient_2017}.
    Although VQE significantly lowered the quantum resource requirement compared to QPE, the circuit depth of an accurate ansatz especially the physically motivated UCC can still be unacceptably large, for example, a naive implement of UCC singles and doubles (UCCSD) ansatz for the 12-qubit LiH molecule introduces $1.6\times 10^{3}$ CNOT gates, which is far beyond the capability of current NISQ devices. Several techniques has been developed to reduce the circuit depth overhead, such as the $k$-UpCCGSD method \textit{et al}\cite{gsd-4} which involves only paired excitations or the qubit-excitation-based method\cite{Yordanov_QEB_2021} which simplifies the Fermionic excitation operators to reduce the CNOT count. However, these methods can potentially lead to worse accuracy than the original UCCSD ansatz. Iterative algorithms such as ADAPT-VQE\cite{adapt-1} which selects operators based on gradients evaluated at each step or iterative qubit-coupled-cluster method\cite{ryabinkin_qubit_2018,  ryabinkin2020iterative, Ryabinkin_iQCC_2021} which dresses the Hamiltonian with optimized Pauli excitations are also proposed to generate a compressed quantum circuit ansatz for a desired accuracy, while the compression is achieved at the expense of increased measurement complexity at each iteration.
	
	In this article, we proposed a circuit- and parameter-efficient method on the basis of the unitary coupled-cluster ansatz. First, the contribution of each cluster operator is measured and sorted on the top of the Hartree-Fock reference (HF) state. The terms that lead to energy reduction over a pre-defined threshold are picked out in order to form a new operator pool. Then, the new pool is used to generate an approximated wave function. Additional operators are added into the new operator pool based on the sorted sequence in the first step and grows the ansatz iteratively. We term the above algorithm as energy-sorting (ES) algorithm. The major advantage of ES-VQE algorithm is that only a single-shot evaluation at the very beginning is used to "score" each operator, and the overhead of measurements will not be significantly increased during the growth of the wave function ansatz. Our benchmark results show that the ES-VQE algorithm is able to retain the accuracy of the original UCCSD in a couple of iterations. With a more robust operator pool, i.e., UCC with generalized singles and doubles (UCCGSD), the ES-VQE algorithm is able to reach chemical accuracy for even strongly correlated system such as H\textsubscript{6} molecule. The number of operators is reduced dramatically after the ES-VQE optimization, thus effectively reduce the depth of quantum circuit. Our algorithm is able to capture the dominant operator terms and greatly reduce quantum resources overhead compare to the original operator pool, making it promising in quantum chemistry simulations on NISQ device. 
	
	The rest parts of the paper are organized as follows. We begin with a brief review of the VQE algorithm and the UCC ansatz in Sec.\ref{sub-vqe}. In Sec.\ref{sub-sort} we give a detailed description of the energy-sorting algorithm. A schematic flowchart is presented to illustrate the implementation of ES-VQE algorithm for quantum chemistry simulations. In Sec.\ref{sec-res} the benchmark results of the ES-VQE algorithm are given for chemical systems including H\textsubscript{4}, LiH, H\textsubscript{6} and the periodic one-dimensional (1D) hydrogen chain. Finally, we conclude our work and provide suggestions for further improvements of the algorithm. 
	
	\section{\label{theory}Method}
	\subsection{\label{sub-vqe}Variational Quantum Eigensolver and Unitary Coupled-Cluster Ansatz}
	The second-quantized formulation of quantum chemistry leads to the Hamiltonian under Born-Oppenheimer approximation as
	\begin{equation}
		\label{eq-ham}
		\hat{H}=\sum_{p,q} {h_{pq} a^{\dagger}_{p} a_{q}} + \sum_{p,q,r,s} {\frac{1}{2} h^{pq}_{rs} a^{\dagger}_{p} a^{\dagger}_{q} a_{r} a_{s}}
	\end{equation}
	where ${a^{\dagger}_{i}}$ and $a_{j}$ are Fermionic creation and annihilation operators, and the one-body and two-body coefficients ${h_{pq}}$ and ${h^{pq}_{rs}}$ in Eq.(\ref{eq-ham}) can be computed on the classical computer.
	The ground-state wave function and energy is then solved from the eigenvalue problem
	\begin{equation}
		\hat{H}|\Psi\rangle = E|\Psi\rangle
	\end{equation}

	In the VQE algorithm, the key ingredient is the parametric unitary operator to prepare the wave function ansatz
	\begin{equation}
		\label{eq-general}
		|\Psi(\vec{\theta})\rangle = U(\vec{\theta})|\Psi_{0}\rangle ,
	\end{equation}
	where the reference wave function $|\Psi_{0}\rangle$ is usually chosen to be the Hartree-Fock state.
	The parametric wave function is then optimized according to Rayleigh-Ritz variational principle
	\begin{equation}
		\label{eq-vqe}
		E = \min_{\vec{\theta}} {\langle\Psi(\vec{\theta})|\hat{H}|\Psi(\vec{\theta})\rangle}
	\end{equation}
	
	In a typical VQE framework, chemical and physical quantities, for example, total energy of the system, are evaluated on a quantum computer. The gradients can be obtain through finite difference steps or using the parameter-shift rule. The optimization of parameters are then performed on a classic computer using conventional optimizer such as Conjugated-Gradient or Simultaneous Perturbation Stochastic Approximation or Nelder–Mead. 
	
	The unitary coupled-cluster\cite{ucc-1, ucc-2, ucc-3} (UCC) ansatz is one of the most commonly used PMA ansatz in electronic structure simulations. Unlike the traditional coupled-cluster theory, the energy and wave function under UCC are determined variationally using Eq.(\ref{eq-vqe}). The unitary operator $U(\vec{\theta})$ is defined as
	\begin{equation}
		\label{eq-ucc}
		|\Psi\rangle = \exp{(\hat{T} - \hat{T}^{\dagger})}|\Psi_{0}\rangle,
	\end{equation}
	where $|\Psi_{0}\rangle$ is chosen to be the single-determinant Hartree-Fock (HF) wave function. The cluster operator that truncated at single- and double-excitations has the form of
	\begin{equation}
		T(\vec{\theta}) = \sum_{p, q}^{\substack{p\in vir\\q\in occ}} {\theta^{p}_{q} \hat{T}^{p}_{q}} + \sum^{\substack{p,q\in vir\\r,s\in occ}}_{\substack{p>q\\r>s}} {\theta^{pq}_{rs} \hat{T}^{pq}_{rs}}
	\end{equation}
	where the one- and two-body terms are defined as
	\begin{equation}
		\hat{T}^{p}_{q} = a^{\dagger}_{p} a_{q}
	\end{equation}
	\begin{equation}
		\hat{T}^{pq}_{rs} = a^{\dagger}_{p} a^{\dagger}_{q} a_{r} a_{s}
	\end{equation}
	Using Fermion-to-Qubit transformations such as Jordan-Wigner or Bravyi-Kitaev\cite{jw-bk-1, jw-bk-2, jw-bk-3, jw-bk-4}, the unitary operator $U(\vec{\theta})=\exp(\hat{T} - \hat{T}^{\dagger})$ can then be written as:
	\begin{equation}
        U(\vec{\theta})
	    = \exp{(i\sum_{p, \alpha}{\tilde{\theta}^{\alpha}_{p} \sigma^{\alpha}_{p}} + i\sum_{pq, \alpha\beta}{\tilde{\theta}^{\alpha\beta}_{pq} \sigma^{\alpha}_{p} \sigma^{\beta}_{q}} + \dots ) }
	\end{equation}
	\begin{equation}
	    \hat{H} = \sum_{p, \alpha}{h^{\alpha}_{p} \sigma^{\alpha}_{p}} + \sum_{pq, \alpha\beta}{h^{\alpha\beta}_{pq} \sigma^{\alpha}_{p} \sigma^{\beta}_{q}} + \dots
	\end{equation}
	where $\sigma^{\alpha}_{p}, \sigma^{ \beta}_{q} \in \{\sigma_{x}, \sigma_{y}, \sigma_{z}, I\}^{\bigotimes}$ and $p, q, \dots$ are indices of qubits, where $\{\tilde{\theta}\}$ and $\{\theta\}$ span the same parameter space.

	On a quantum computer, the implementation of the VQE circuit for UCCSD ansatz requires decomposition of the exponential-formed cluster operators into basic quantum single-qubit and two-qubit gates. Approximation schemes are often used, such as Trotter-Suzuki decomposition\cite{trotter-1, trotter-2}:
	\begin{equation}
		\exp{(\hat{A}+\hat{B})} =  \lim_{N\rightarrow\infty}{(e^{(\hat{A}/N)} e^{(\hat{B}/N)})^{N}}
	\end{equation}
	The Trotterized UCC wave function takes the form:
	\begin{equation}
		|\Psi\rangle = {\prod^{N}_{k=1}{\prod^{N_{i}}_{i}{e^{\frac{\theta_{i}}{N}\hat{\tau}_{i}}}}}|\Psi_{0}\rangle,
	\end{equation}
	where $N_{i}$ is the total number of individual operators $\hat{\tau}_{i}$.
	The Trotterization is truncated at finite order $N$, and the number of Trotter steps, the ordering sequence of operators and the Trotter formula used in the procedure will have significant influence on the accuracy of the simulation. In this study, we use the single-step Trotter decomposition. Here, we summarize the procedure of the VQE method with unitary coupled-cluster ansatz (UCC-VQE) in \textbf{Algorithm \ref{algo1}}
	
	\begin{algorithm}[H]
		\label{algo1}
		\caption{Procedure of VQE algorithm for UCC ansatz}
		\SetAlgoLined
		\KwIn{\\\ \ Reference state $|\Psi_{0}\rangle$(such as $| \Psi_{HF} \rangle$)\\\ \ System Hamiltonian $\hat{H}$}
		\KwOut{\\\ \ The optimized ansatz and energy}
		Prepare initial wave function\;
		Generate parametric cluster operators and map the unitary cluster operators to quantum circuit; 
		\While{not converged, i.e. $\Delta E^{(k)} \ge \varepsilon$}
		{
			Measure multiple times to obtain the expectation of $E^{(k)}=\langle \Psi(\vec{\theta}^{(k)}) | \hat{H} | \Psi(\vec{\theta}^{(k)}) \rangle$
			\;
			Perform classical optimization algorithm to update parameters: $\vec{\theta}^{(k)}\rightarrow \vec{\theta}^{(k+1)}$\;
			Update quantum circuit for the next iteration
		}
	\end{algorithm}
	
	\subsection{\label{sub-sort}Energy-sorting VQE algorithm}
	
	\begin{figure}
		\begin{center}
			\includegraphics[width=0.4\textwidth]{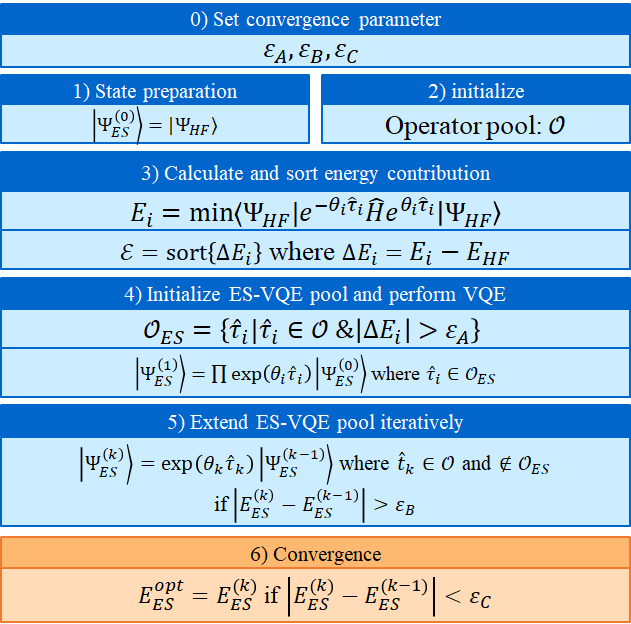}
			\caption{Procedure of the ES-VQE scheme. The energy contribution $\mathcal{E}=\{\Delta E_{i}\}$ of each operator $\hat{\tau}_{i} \in \mathcal{O}$ is calculated and sorted through a single-parameter VQE circuit. At the 1-st step, the ES-VQE pool is initialized by all operators $\{\hat{\tau}_{i}\}$ for which $\Delta E_{i} > \varepsilon_{A}$. At the $k$-th ($k>1$) ES-VQE iteration, operator $\hat{\tau}_{k}$ which is in $\mathcal{O}$ but not in $\mathcal{O}_{ES}$ is selected, such that $|E^{(k)}_{ES} - E^{(k-1)}_{ES}| > \varepsilon_{B}$. The convergence is achieved at the $k$-th iteration if $|E^{(k)}_{ES} - E^{(k-1)}_{ES}| < \varepsilon_{C}$}.
			\label{fig1algo2}
		\end{center}
	\end{figure}
	
	VQE significantly relieves the hardware demand compared to QPE and the trading in long circuit depth is achieved at the expense of a much higher number of measurements. Our proposed energy-sorting algorithm would require circuits with much shallower depth and reduced number of measurements dramatically, by selecting only the important terms in the original operator pool. We term this protocol as the ES-VQE method.
	
	As a typical post-Hartree-Fock method, the coupled-cluster (CC) wave function is treated as an exponential of the cluster operators on a reference state:
	\begin{equation}
		\label{eq-cc}
		| \Psi \rangle = \exp{(\hat{T})} | \Psi_{0} \rangle			
	\end{equation}
	Expanding the exponential formalism, we can simply write the CC energy as:
	\begin{equation}
	    \begin{aligned}
	        E &= \langle \Psi_{0} e^{-\hat{T}} | \hat{H} | e^{\hat{T}} \Psi_{0} \rangle \\
	        & \approx \langle \Psi_{0} | \hat{H} | \Psi_{0} \rangle + \sum_{i, j}^{}{t_{ij}\langle \Psi_{i} | \hat{H} | \Psi_{j} \rangle}
	    \end{aligned}
	\end{equation}
	Due to the exponential ansatz, CC can implicitly include higher excitation operators, leading to fast convergence behaviour. For UCCSD, a similar expansion of energy can be obtained by replacing $\hat{T}$ with $\hat{T} - \hat{T}^{\dagger}$. In general, the total energy can be expressed as the Hartree-Fock energy $E_{0}=\langle \Psi_{0} | \hat{H} | \Psi_{0} \rangle$ plus the correlation part $\sum{t_{ij}\langle \Psi_{i} | \hat{H} | \Psi_{j} \rangle}$. Each cluster operator contributes a part of the correlation energy. However, their contributions are not necessarily equal, for example, excitations near the highest occupied orbitals (HOMO) or lowest unoccupied orbitals (LUMO) may have larger contributions. Based on this observation, a compact wave function ansatz based on UCCSD can be constructed by adding operators according to their "importance". Here we used the contribution to the total energy of an operator as an indicator of its "importance".
		
	The basic outline of ES-VQE is drawn schematically in Fig.(\ref{fig1algo2}). Here we described the ES-VQE method as follows:
	\begin{itemize}
        \item [0)]
        Set the input parameters for ES-VQE: $\varepsilon_{A}$, $\varepsilon_{B}$ and $\varepsilon_{C}$.
		\item [1)]
		Initialize the qubits or quantum register, and prepare the reference state. In this work, we use the Hartree-Fock state as the reference state. The quantities such as one- and two-electron integrals for preparing the initial input is calculated on a classical computer.
		\item [2)]
		Define an operator pool $\mathcal{O}$ to build the wave function ansatz, for example, UCCSD or UCCGSD operator pool. The operator pool contains all the terms which can be used to generate the wave function ansatz.
		\item [3)]
		On a quantum computer, perform VQE optimization iterations for each operator $\hat{\tau}_{i} \in \mathcal{O}$. This can be carried out in parallel if multiple quantum processors are available. Evaluate the energy $E_{i}$ optimized using only a single cluster operator $\hat{\tau}_{i}$ and calculate its difference relative to the reference state energy: $\Delta E_{i} = E_{i}-E_{HF}$. Store the $\{\Delta E_{i}\}$ as a sorted list $\mathcal{E}=\text{sort}\{(\Delta E_{i}, \hat{\tau}_{i})\}$ in descending order.
		\item [4)]
		The first iteration starts with a subset of operators from the original pool. Select operators $\{\hat{\tau}_{i}\}$ with total energy contribution above a threshold $\Delta E_{i} > \varepsilon_{A}$. Perform VQE to optimized parameters in the generated ansatz.
		\item [5)]
		Select the next one operator $\tau_{k}$ which is in the original pool $\mathcal{O}$ but not yet been used according to the sequence in the sorted list ${\mathcal{E}}$. The wave function ansatz is updated using the selected terms on top of the previous chosen operators. VQE is performed again and re-optimize all the parameters. If the absolute difference in energy of the $k$-th and ($k-1$)-th iteration is larger than $\varepsilon_{B}$, accept the ansatz as ${|\Psi^{(k)}_{ES}}\rangle = \exp(\theta_{k} \tau_{k}) {|\Psi^{(k-1)}_{ES}}\rangle$. 
		\item[6)]
		If the convergence criteria $\varepsilon_{C}$ is satisfied, i.e., $|E^{(k)}_{ES} - E^{(k-1)}_{ES}| < \varepsilon_{C}$, then output the optimized wave function $|\Psi^{opt}_{ES}\rangle$ together with corresponding energy $E^{opt}_{ES}$ and exit step 5.
		
	\end{itemize}	
	
	As described above and in Fig.(\ref{fig1algo2}), at the beginning of ES-VQE algorithm, an additional loop of VQE will be performed to estimate the influence of each operator on the total energy:
	\begin{equation}
		E_{i}=\min_{\theta_{i}} {\langle \Psi_{HF} | e^{-\theta_{i} \hat{\tau}_{i}} \hat{H} e^{\theta_{i} \hat{\tau}_{i}} | \Psi_{HF} \rangle}
	\end{equation}
	The result $\Delta E_{i} = E_{i} - E_{HF}$ is then stored and sorted in a list, namely $\mathcal{E}$ on the classical computer. Although the gradient can also be used as a "score" for each operator, directly calculating energy contribution is more precious as pointed out by Armaos\textit{et al.}\cite{Armaos2021e_vs_g}.  The next iterations will take the excitation operators following the order in $\mathcal{E}$. At the first iteration, all the operators ${\hat{\tau}_{i}}$ with $\Delta E_{i}$ larger than a given threshold $\varepsilon_{A}$ will be used to construct a wave function ansatz which is an approximation to the unknown ground state. 
	After this step, extra iterations are carried which is described in step 5. Starting from the 2$nd$ iteration, the ansatz is growing by adding one term corresponding to the next largest $\Delta E_{i}$ in the sorted list $\mathcal{E}$ and re-optimizing all the parameters. For the operators which do not have contributions on a Hartree-Fock state, the order is kept the same as that they were generated in step 2. Note that for different states, for example, ${|\Psi^{(k)}_{ES} \rangle}$ and ${|\Psi_{HF} \rangle}$, the contributions of a single operator ${\hat{\tau}_{i}}$ to the total energy in the Trotterized wave function may not be the same, therefore the second threshold parameter $\varepsilon_{B}$ is necessary to ensure that all of the operators used in the final ansatz have non-zero amplitudes. This step will terminate if the energy different of two consecutive steps is smaller than $\varepsilon_{C}$. Notice that setting $\varepsilon_{B} > \varepsilon_{C}$ is equivalent to terminating the ES-VQE iteration only if all the operators in $\mathcal{O}$ has been tested in Step 5). Additional criteria can also be introduced, for example whether $E^{(k)}_{ES}$ is below some pre-estimated threshold.
	

	\section{\label{sec-res}Results}
	We demonstrate our method for molecular and periodic systems. The benchmarks are performed using a in-house developed code to perform the ES-VQE algorithm. The one- and two-body integrals are calculated using the PySCF software\cite{pyscf}. The mapping from Fermion operators to qubit representations are obtained by Jordan-Wigner transformation using MindQuantum\cite{MindQuantum}. The gradient-based optimization is performed utilizing the Broyden-Fletcher-Goldfarb-Shannon (BFGS) algorithm which is implemented in the SciPy package\cite{scipy} with convergence criteria gtol=$10^{-7}$ a.u. The wave function ansatz constructed through UCC-VQE and ES-VQE involves single-step Trotter-Suzuki decomposition. The ES-VQE parameters are set as $\varepsilon_{A}=\varepsilon_{B}=1\times 10^{-4}$ and $\varepsilon_{C} = 1\times 10^{-8}$. We put a restriction in step 5 of Fig.(\ref{fig1algo2}) that the operators from the original operator pool can only be tested once. In addition, for benchmark purpose, we calculated FCI energies as a reference and terminates the ES-VQE iteration if $|E_{ES} - E_{FCI}| \le 1.0$ kcal/mol.

\begin{table}[]
	    \centering
	    \begin{tabular}{cccccc}
\hline 
 & UCCSD & UCCGSD & QCC & sym-UCCSD & sym-UCCGSD \\
\hline
H$_4$ & 14 & 72 & 848 & 8 & 32
 \\
LiH & 44 & 345 & 4984 & 20 & 117
 \\
H$_6$ & 54 & 345 & 4984 & 29 & 165
\\
H$_2$-PBC & 36 & 240 & 4984 & / & /
\\

\hline \hline 
	         
	    \end{tabular}
	    \caption{Number of operators using different ansatzes for H$_4$, H$_6$, LiH and the hydrogen chain. The QCC operator pool is obtained by generating Pauli strings containing no Pauli $Z$ and odd number of Pauli $Y$, with a maximum length of 4.}
	    \label{tab::h4_h6_h8}
	\end{table}
 
	\begin{figure*}
		\subfigure[]
		{
			\includegraphics[width=0.3\textwidth]{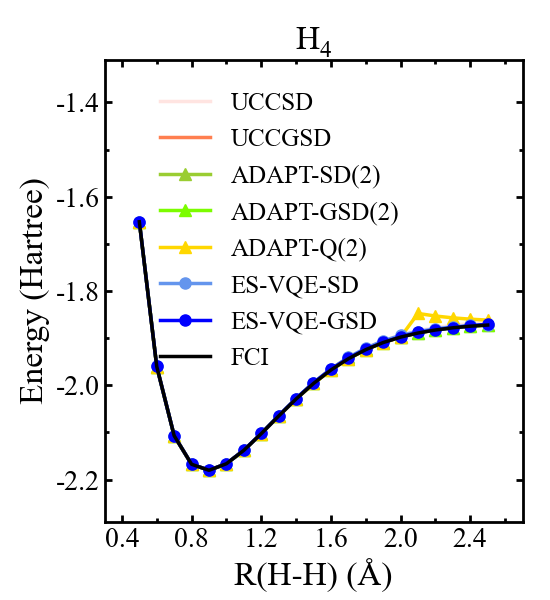}
		}
		\subfigure[]
		{
			\includegraphics[width=0.3\textwidth]{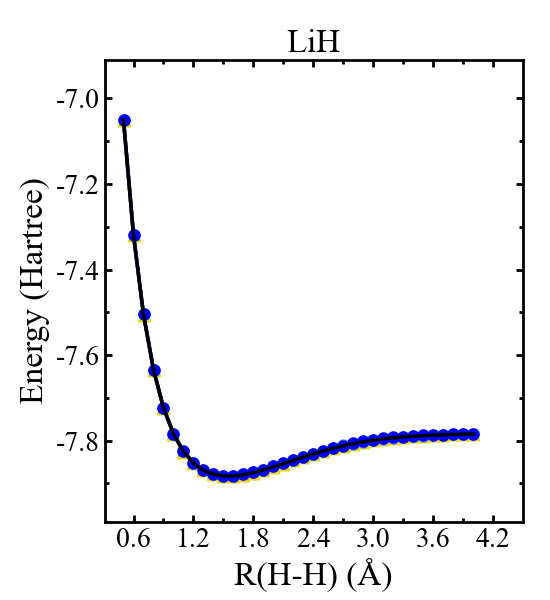}
		}
		\subfigure[]
		{
			\includegraphics[width=0.3\textwidth]{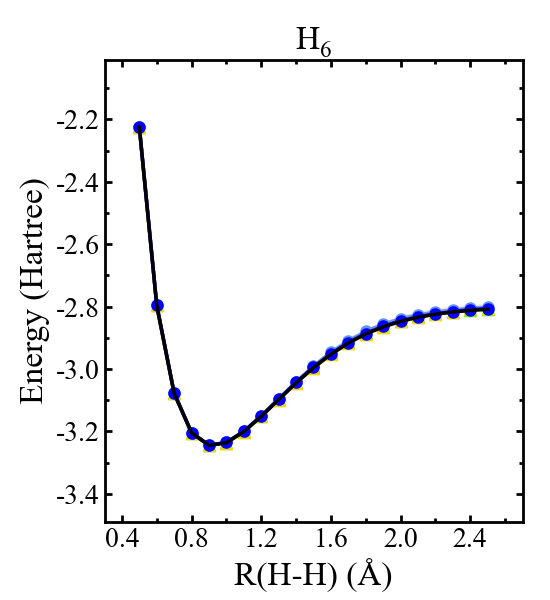}
		}
		\subfigure[]
		{
			\includegraphics[width=0.3\textwidth]{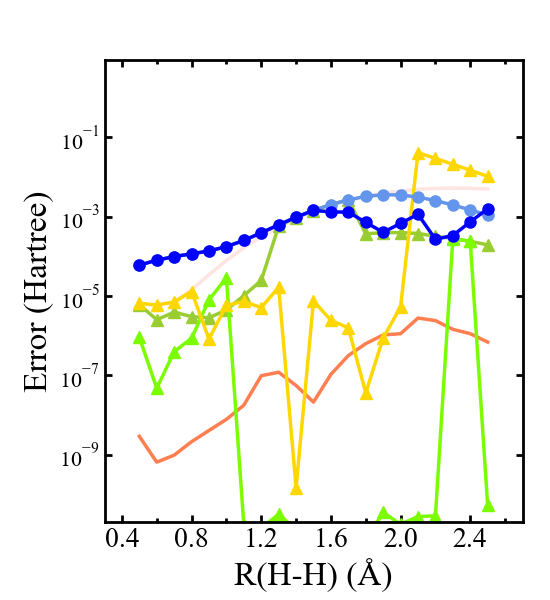}
		}
		\subfigure[]
		{
			\includegraphics[width=0.3\textwidth]{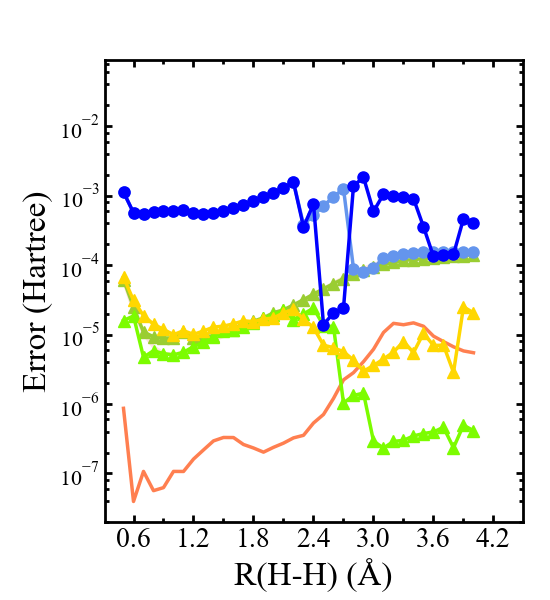}
		}
		\subfigure[]
		{
			\includegraphics[width=0.3\textwidth]{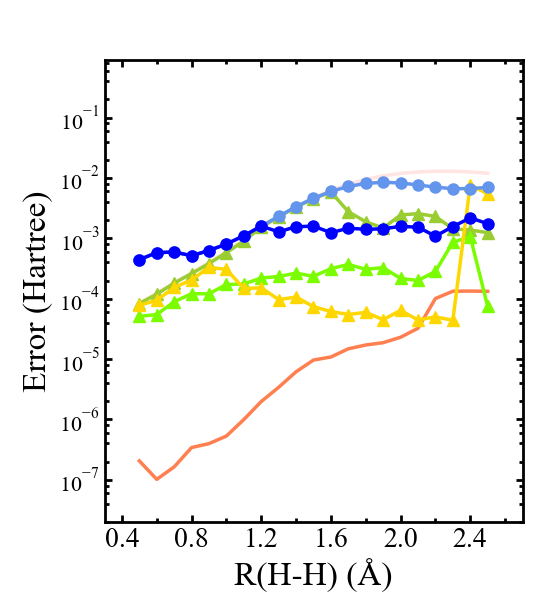}
		}
		\subfigure[]
		{
			\includegraphics[width=0.3\textwidth]{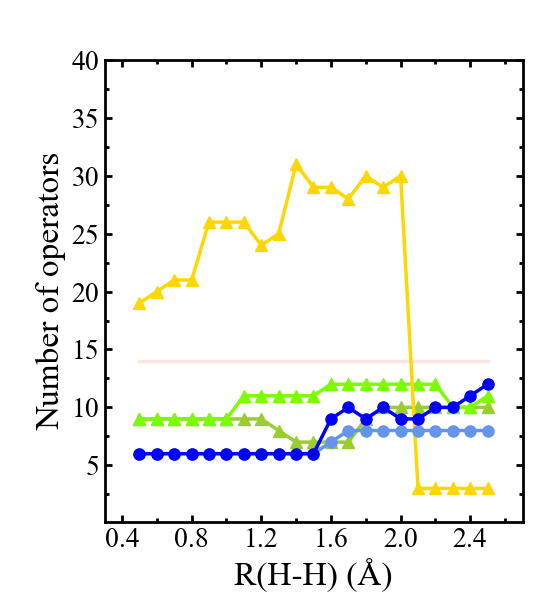}
		}
		\subfigure[]
		{
			\includegraphics[width=0.3\textwidth]{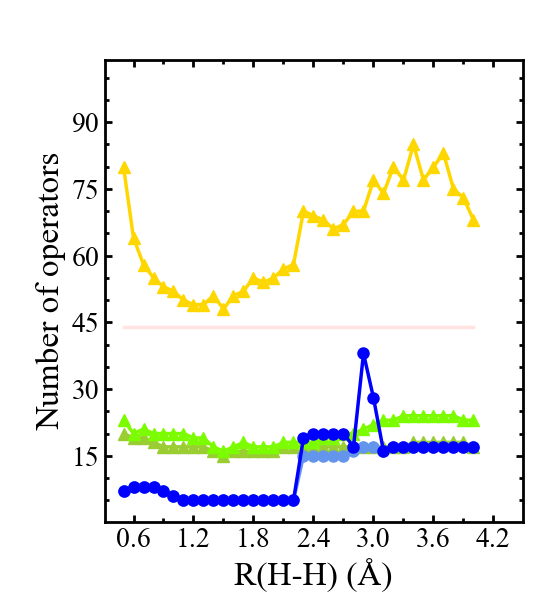}
		}
		\subfigure[]
		{
			\includegraphics[width=0.3\textwidth]{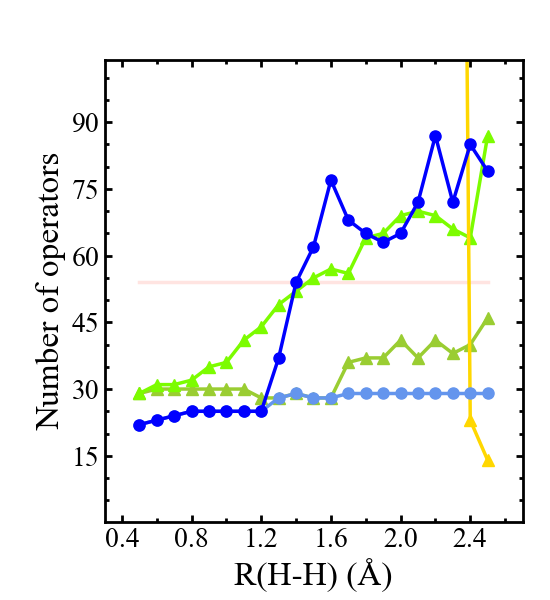}
		}
		\caption{The calculated results for H\textsubscript{4}, LiH and H\textsubscript{6}. (a,b,c). Potential energy surface calculated using ES-VQE-SD/ES-VQE-GSD in atomic units (Hartree). (d,e,f). Absolute error in energy of different method with respect to FCI, unit in Hartree. The energy error is plotted using logarithmic axis. (g,h,i). Number of operators in the final ansatz optimized by ES-VQE-SD/ES-VQE-GSD. Data points where the number of operators exceeds 105 (such as UCCGSD and part of ADAPT-Q) are not displayed in (i). Note that a large number of operators for ADAPT-Q does not necessarily indicate large circuit depth.}
		\label{fig2}
	\end{figure*}

	For molecular systems, three molecules H\textsubscript{4}, LiH and H\textsubscript{6} are tested. The minimum STO-3G basis set is used for the calculations. The calculated potential energy surface as a function of nuclear coordinates are shown in Fig.(\ref{fig2}), together with error with respect to the FCI energy and the number of operators used in the final wave function ansatz. Results of ES-VQE using UCCSD and UCCGSD operator pool (termed as ES-VQE-SD and ES-VQE-GSD) are benchmarked against UCCSD, UCCGSD and ADAPT-VQE. For the ADAPT-VQE calculations, we tested the UCCSD (ADAPT-SD), UCCGSD (ADAPT-GSD) and the qubit coupled-cluster (QCC)\cite{ryabinkin_qubit_2018} operator pool (ADAPT-Q) which includes Pauli strings with a maximum lengh of 4. The ADAPT-Q is also recognized as the qubit-ADAPT-VQE\cite{Tang2021qubitadapt}. The convergence threshold of ADAPT-VQE is set as $10^{-2}$ for residual gradients, which is denoted as ADAPT-X(2) where X$\in \{SD, GSD, Q\}$.

    Table~\ref{tab::h4_h6_h8} shows the size of the operator pool of different ansatzes without iterative optimization. In general, ES-VQE-SD successfully maintains the accuracy comparable to original UCCSD using the complete operator pool, meanwhile significantly reduces the number of variational parameters with a comparable or even greater factor than the symmetry reduced UCC\cite{cao_SYMM_2022}. Notice that For H$_4$ and H$_6$ at large bond length, the ES-VQE-SD performs even better thant the original UCCSD, indicating that a better Trotter sequence is generated by ES-VQE. If the UCCGSD operator pool is introduced (the size of which is 72 for H$_4$ and 345 for LiH and H$_6$), chemical accuracy (error within 1.0 kcal/mol, i.e., approximately $1.6\times 10^{-3}$ Hartree with respect to the FCI energy) is achieved throughout the potential energy curve for all three tested molecules using ES-VQE-GSD.  For H\textsubscript{4} approximately 50\% of the operators are removed using ES-VQE-SD, and this number is increased to $\sim$85\% for ES-VQE-GSD. For H\textsubscript{6} which have more significant strongly-correlated effect, a reduction of 50\% to 93\% is observed, depending on the bond length. While for weakly-correlated systems such as LiH, a reduction factor up to 98\% is obtained (ES-VQE-GSD), especially at shorter Li-H distance.
    
    Since the contribution of each operator is calculated based on the Hartree-Fock state which is not a good reference state in strongly correlated systems, the step 4 in Fig.(\ref{fig1algo2}) becomes inaccurate, leading to the size of the ES-VQE pool increasing rapidly and the performance of ES-VQE can become worse than ADAPT-X(2) with even more operators, as shown in the results for H$_4$ and H$_6$ at large bond length. Nevertheless, the ES-VQE is suitable for weakly correlated systems, and the one-shot evaluation of operator contributions brings a smaller complexity than ADAPT-VQE in subsequent iterations. It should be noted that, for the ADAPT-Q Pauli strings instead of fermionic excitation operators are used. Due to the small length of Pauli strings, a large number of operators does not necessarily lead to a great circuit depth. However, the measurement overhead will still be remarkable due to the evaluation of gradients of a large number of operators at each iteration. 

\begin{figure*}[ht!]
		\subfigure[]
		{
			\includegraphics[width=0.3\textwidth]{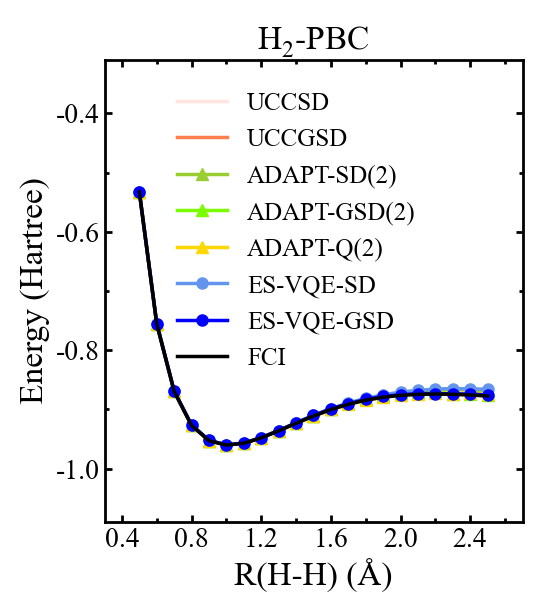}
		}
		\subfigure[]
		{
			\includegraphics[width=0.3\textwidth]{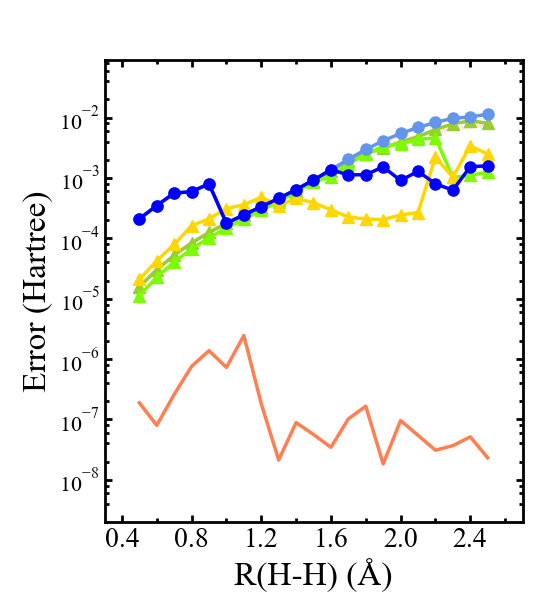}
		}
		\subfigure[]
		{
			\includegraphics[width=0.3\textwidth]{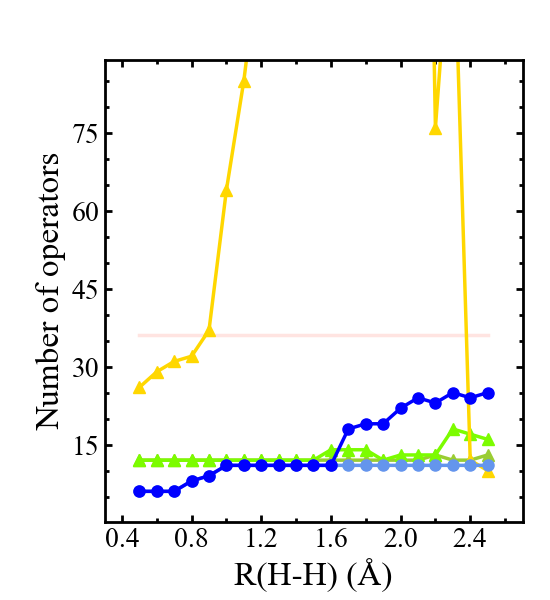}
		}
		\caption{The benchmark results of ES-VQE-SD and ES-VQE-GSD for 1D equi-spaced hydrogen chain. Two atoms per primitive cell with three sampled $k$-points are used in the calculation. An extended operator pool with auxiliary terms defined by Eq.(\ref{res-term1})-(\ref{res-term2}) is used in order to minimize the residual error from complex wave functions. (a). Potential energy curve in unit Hartree of 1D hydrogen chain as a function of H-H distance. (b). Deviation of energy calculated by different methods from FCI results in logarithmic axis. (c). Total number of terms in the final ansatz. Part of the curve for ADAPT-Q where the number of operators exceeds 90 is not shown.}
		\label{fig3}
	\end{figure*}

    We also test our algorithm for the periodic one-dimensional hydrogen chain. We use the SVZ basis set together with GTH pseudopotential\cite{gth-pp}. The unit cell consists of two hydrogen atoms. Three $k$-points are sampled along the chain, making a total number of 12 qubits in the simulation. Quantum simulations for periodic systems can be challenging and may faced with both accuracy and resource issues due to the residual error introduced by complex wave function and the necessity for dense $k$-point sampling grid. Our previous work\cite{pbc-1, FAN_2022_PBC} pointed out that the original implementation of the unitary coupled-cluster method failed to give accurate prediction of ground states in periodic systems, and a complementary operator pool is necessary to achieve accurate ground state energy:
	\begin{equation}
		\label{res-term1}
		\tilde{\tau}^{p}_{q} = i(a^{\dagger}_{p} a_{q} + a^{\dagger}_{q} a_{p})
	\end{equation}
	\begin{equation}
		\label{res-term2}
		\tilde{\tau}^{pq}_{rs} = i(a^{\dagger}_{p} a^{\dagger}_{q} a_{r} a_{s} + a^{\dagger}_{s} a^{\dagger}_{r} a_{q} a_{p})
	\end{equation}
	More details about the residual error is presented in Appendix~\ref{AppACSE}.

	The benchmark results for 1D hydrogen chain are presented in Fig.(\ref{fig3}). The FCI results used for benchmark are obtained by directly diagonalizing the Hamiltonian. Similar to H\textsubscript{4} and H\textsubscript{6}, although ES-VQE-SD successfully reaches the chemical accuracy of UCCSD, the UCCSD itself is not able to provide chemical accuracy across the potential energy surface with all the single- and double-excitation operators traversed. We observed that using the UCCGSD operator pool, ES-VQE-GSD greatly suppressed the error to below chemical accuracy at the expense of a few more operators, which is in agreement with previous studies that UCCGSD provides more stable and accurate ground-state predictions not only in molecular calculations but also in periodic systems\cite{gsd-4, pbc-1}.
	By identifying and gathering the dominant excitations which have large contributions to total energy, ES-VQE-SD algorithm shows excellent performance for the 1D hydrogen chain, by selecting 20$\sim$50 percent of the operators from the original UCCSD pool. As for ES-VQE-GSD, although more terms are required to reach chemical accuracy at larger H-H distance, the reduction in the number of operators is still remarkable, with a factor of 89\%$\sim$97\% (6$\sim$25 out of 240 UCCGSD operators). Applications in larger periodic systems are therefore promising.

	It is worthy noting that at step 5 of the ES-VQE algorithm, the number of parameters to be re-optimized contains all $n_{k+1}$ terms, where $n_{k+1}$ is the number of operators used at the $(k+1)th$ iteration. 
	We note that for the same term $\hat{\tau}_{i_{0}}$, the optimized value of its coefficient between different iterations $k_{1}$ and $k_{2}$ are generally different. An alternative optimization scheme is tested by freezing the previously optimized $n_{k}$ parameters while only optimized the newly added term. Unfortunately, deterioration in both number of operators and accuracy is observed. The results are shown in Appendix~\ref{Appopt}.
	
	\section{\label{conclu}Conclusion}
	In this work, we proposed an optimized energy-sorting algorithm named ES-VQE to build the wave function ansatz. With this algorithm, effective excitation operators are identified to form an ordered sequence. The wave function ansatz is then constructed using this in-order operator pool. ES-VQE successfully achieved UCCSD accuracy for both non-periodic and periodic systems, with a significant reduction in term of operator pool size, thus effectively reduces the quantum circuit depth and the number of measurements. By using a more robust operator pool such as UCCGSD\cite{gsd-4, pbc-1}, ES-VQE shows the ability to achieve higher precision of the ground-state energy at the expense of a few more operators. Our algorithm shows a potential of huge resource reduction while preserving the simulation accuracy, paving the way toward efficiently simulating larger chemical systems on near-term quantum computers. 
	
	To make the ES-VQE algorithm more suitable for the NISQ device, quantum circuit compilation and optimization that consider hardware constraints is needed to further suppress the circuit depth. Besides, efficient measurement scheme is also required in order to reduce the overhead of measurements. By combining the ES-VQE algorithm with other methods such as parameter reduction techniques considering point-group symmetry, efficient large-scale simulations for both molecular and periodic systems can be performed on NISQ devices.
	
	 
	\appendix

	\section{\label{AppESalgo}Identifying the effective operators}

    The wave function ansatz used in VQE can be expressed using a general form:
    \begin{equation}
        | \Psi (\vec{\theta}) \rangle = \prod_{i} {U_{i} (\theta_{i})} | \Psi_{HF} \rangle,
    \end{equation}
    where $\{U_{i}\}$ can be operators which have physical meanings such as Fermionic excitation operators, or simply parametric quantum gates. In most cases, the parameter space $\{\theta_{i}\}$ is redundant, and only a subset of $\{U_{i}\}$ is needed to reach a given accuracy. Therefore, identifying the effective $U_{i}$s is a central part to reduce the circuit depth. A straightforward strategy is to "score" each unitary, which gives an estimation of how important the unitary is in contributing to the accuracy of the VQE energy. Using this strategy, not only the number of parameters but also the sequence of $\{U_{i}\}$ can be optimized.
    
    For example, the ADAPT-VQE algorithm uses gradients as the score for each unitary:
    \begin{equation}
        s^{(k)}_{i} = \frac{\partial{\langle \Psi^{(k-1)} U^{\dagger}_{i}(\theta_{i}) | \hat{H} | U_{i}(\theta_{i}) \Psi^{(k-1)} \rangle}}{\partial \theta_{i}} \bigg|_{\theta_{i}=\theta_{i, 0}},
    \end{equation}
    where $s^{(k)}_{i}$ is the score of unitary $U_{i}(\theta_{i})$ at the $k$-th iteration. However, the largest gradient does not necessarily indicate the largest contribution to the total energy, and evaluating all the $\{s^{(k)}_{i}\}$ at each iteration will increase the overall complexity from $\mathcal{O}(N_{q}^4)$ to $\mathcal{O}(N_{q}^4\times N_{p} \times K)$ if $K$ iterations are performed, where $N_{q}$ is the number of qubits and $N_{p}$ is the number of parameters.
    In this study, we use the energy difference as the score:
    \begin{equation}
	    \begin{aligned}
	        s^{(k)}_{i} &= \langle \Psi^{(k-1)} U^{\dagger}_{i}(\theta_{i,opt}) | \hat{H} | U_{i}(\theta_{i,opt}) \Psi^{(k-1)} \rangle \\
                    &- \langle \Psi^{(k-1)} | \hat{H} | \Psi^{(k-1)} \rangle.
	    \end{aligned}
	\end{equation}
    In addition, we let $s^{(k)}_{i}=s^{(0)}_{i}$ for $k\ge 1$. In this way, the expensive scoring procedure is carried only once at the $0$-th iteration, and the measurement overhead is still at the order of $\mathcal{O}(N_{q}^4)$ in subsequent iterations. However, the precision of such a one-shot score is system specific. Since we use the Hartree-Fock wave function as a initial state, it is expected that this method can have unsatisfying performance for strongly correlated systems as presented in the benchmark results.

	\section{\label{AppACSE}Residual error in the UCC ansatz for periodic systems}
	The anti-Hermitian contracted Schr\"{o}dinger equation described the convergence criteria as\cite{acse-1, pbc-1}:
	\begin{equation}
		\label{eq-CSE}
		\langle \Psi |[ \hat{T}_{u}, \hat{H} ]| \Psi \rangle = 0
	\end{equation} 
	where $\hat{T}_{u}$ is a general two-body operator. Eq.(\ref{eq-CSE}) can be separated into the summation of real (ACSE-Re) and imaginary (ACSE-Im) part:
	\begin{equation}
		\label{eq-acse-re}
		\langle \Psi |[ \hat{T}_{u} - \hat{T}^{\dagger}_{u}, \hat{H} ]| \Psi \rangle = 0
	\end{equation}
	\begin{equation}
		\label{eq-acse-im}
		\langle \Psi |[ \hat{T}_{u} + \hat{T}^{\dagger}_{u}, \hat{H} ]| \Psi \rangle = 0
	\end{equation}
	The convergence condition of Trotterized UCC-VQE with real-valued parameters is equivalent to the real part Eq.(\ref{eq-acse-re}) of the anti-Hermitian CSE. For the imaginary part, although it is not directly optimized during VQE, non-periodic systems with real-valued wave function ensures its value to be constantly zero.
	
	As the complex-valued wave function is used for periodic systems, the residual error, which refers to the imaginary part of anti-Hermitian CSE in Eq.(\ref{eq-acse-im}), will have non-zero values and not guarantee to be minimized through VQE procedure. Our previous study\cite{pbc-1} shows that in the final ansatz, the residual error can be larger than $8\times10^{-3}$ Hartree. One solution is to use the K2G transformation\cite{pbc-1}, which transforms the complex HF orbitals to real-valued wave function for a $\Gamma$-centered supercell. Another solution employs complex-valued variational parameters and separately optimizing the real and imaginary parts, as studied by Manrique \textit{et al}.\cite{pbc-2}. We used a more flexible method\cite{FAN_2022_PBC} by adding the terms in Eq.(\ref{eq-acse-im}) back into the unitary coupled-cluster ansatz as a auxiliary operator pool, shown as Eq.(\ref{res-term1}) and Eq.(\ref{res-term2}).
	
	\section{\label{Appopt}Optimization scheme}
	
	An alternative approach to perform VQE optimization is to treat the wave function ansatz at $(k+1)th$ iteration as:
	\begin{equation}
		| \Psi^{(k+1)} \rangle = \exp{(\theta^{(k+1)}_{n_{k+1}} \hat{\tau}_{n_{k+1}})} \prod_{i}^{n_{k}} {\exp{(\theta_{i}^{(k)} \hat{\tau}_{i})}} | \Psi_{HF} \rangle
	\end{equation}
	where $\{\theta_{i} | i=1\cdots n_{k}\}$ is the optimized parameters of the previous iteration. The original optimization scheme, namely full-parameter optimization, optimizes all $n_{k+1}$ parameters simultaneously. However, in the single-parameter optimization strategy, the first $n_{k}$ amplitudes will be kept fixed during the VQE procedure. In order to show the difference between these two methods clearly, we choose the periodic 1D hydrogen chain with distance between two hydrogen atoms in the unit cell R(H-H)=2.0 Å as an example. The UCCGSD operator pool is used here to minimize the possible error introduced by the UCCSD pool. The convergence criteria is unchanged, as in the main text.
	
	\begin{table}[h]
		\begin{ruledtabular}
			\begin{tabular}{ccc}
				{ES-VQE-GSD}&$\Delta E_{abs}$ (Hartree)&$N_{ops}$ \\ \hline
				{Full-parameter} & $9.09\times 10^{-4}$ & 22 \\
				\tabincell{c}{Single-parameter} & $3.63\times 10^{-3}$ & 240 \\
			\end{tabular}
		\end{ruledtabular}
		\caption{Results of different optimization schemes used in step 5 of ES-VQE. For full-parameter optimization, all $n_{k+1}$ parameters are optimized simultaneously at the $(k+1)th$ iteration. Single-parameter strategy only update the coefficient of the newly added term while frozen other values. The test is performed using 1D hydrogen chain at R(H-H)=2.0 Å and UCCGSD supplied operator pool.}
		\label{tab1}
	\end{table}

	Table~\ref{tab1} listed the results of two different optimization strategies. Using the UCCGSD operator pool, ES-VQE-GSD successfully reaches chemical accuracy with 20 operators selected from 240 operators. However, for the single-parameter update scheme, we fail to obtain the ground state 
    energy with chemical accuracy even with all the 240 operators. This is due to the fact that single-parameter update scheme limits the optimization to a tiny region near the previous state. Therefore, updating only the last parameter will possibly be unable to find the correct minimum in the whole parameter space spanned by $\{\theta^{(k+1)}_{i}\ |\ i=1\dots n_{k+1}\}$.
	
	\section{\label{Appkpt}$k$-point sampling scheme and periodic Hartree Fock theory}
	In calculations of periodic systems, a finite number of $k$-points are required to sample the Brillouin zone. In this work, we used a gamma-centered Monkhorst-Pack scheme\cite{mp-scheme} for generating $k$-points for the 1D equi-spaced hydrogen chain. A Monkhorst-Pack grid is a rectangular grid of points sampled evenly throughout the first Brillouin zone with dimension $N_{k_{x}} \times N_{k_{y}} \times N_{k_{z}}$. 
	
	Take the 1D hydrogen chain with H-H distance 1.0 Å and three sampled $k$-points as an example:
	\begin{itemize}
		\item [1)]
		The unit cell can be represented by a 3-dimensional matrix:
		\begin{equation}
			R = \left[\begin{matrix}\bm{a_{1}} \\ \bm{a_{2}} \\ \bm{a_{3}} \end{matrix}\right] = \left[ 
			\begin{matrix}
				10.0 & 0.0 & 0.0 \\
				0.0 & 10.0 & 0.0 \\
				0.0 & 0.0 & 2.0
			\end{matrix}
			 \right]
		\end{equation}
		Two hydrogen atoms are aligned along $z$-axis (the direction of $\bm{a_{3}}$). 
		\item [2)]
		The reciprocal lattice vectors $\{\bm{b_{1}}, \bm{b_{2}}, \bm{b_{3}}\}$ can be calculated using:
		\begin{align}
			\label{rec_latt1}
			\bm{b_1} &= 2\pi \frac{\bm{a_2} \times \bm{a_3}}{\bm{a_1} \cdot (\bm{a_2} \times \bm{a_3})} \\
			\bm{b_2} &= 2\pi \frac{\bm{a_3} \times \bm{a_1}}{\bm{a_2} \cdot (\bm{a_3} \times \bm{a_1})} \\
			\bm{b_3} &= 2\pi \frac{\bm{a_1} \times \bm{a_2}}{\bm{a_3} \cdot (\bm{a_1} \times \bm{a_2})}
		\end{align}
		which is equivalent to
		\begin{equation}
			\label{rec_latt2}
			G = \left[\begin{matrix}\bm{b_{1}} \\ \bm{b_{2}} \\ \bm{b_{3}} \end{matrix}\right] = 2\pi \cdot (R^{T})^{-1} 
		\end{equation}
		Therefore, the reciprocal lattice vectors in this example is obtained using Eq.(\ref{rec_latt1})-(\ref{rec_latt2}) (Å\textsuperscript{-1}):
		\begin{equation}
		G = \left[ 
			\begin{matrix}
				0.6283 & 0.0 & 0.0 \\
				0.0 & 0.6283 & 0.0 \\
				0.0 & 0.0 & 3.1416
			\end{matrix}
			\right]
		\end{equation}
    	\item [3)]
    	For a $1\times1\times3$ grid, the coordinates for the sampled $k$-points are thus (Å\textsuperscript{-1}):
    	\begin{align}
    		\label{kpt1}
    		\vec{K}_{1} &= \frac{0}{3} \cdot \bm{b_{3}} = \left(\begin{matrix}0.0, & 0.0, & 0.0 \end{matrix}\right) \\
    		\vec{K}_{2} &= \frac{1}{3} \cdot \bm{b_{3}} = \left(\begin{matrix}0.0, & 0.0, & 1.0472 \end{matrix}\right) \\
    		\vec{K}_{3} &= -\frac{1}{3} \cdot \bm{b_{3}} = \left(\begin{matrix}0.0, & 0.0, & -1.0472 \end{matrix}\right) 
    	\end{align}
	\end{itemize}
	
	Bloch atomic orbitals are then defined as:
	\begin{equation}
	    \chi_{\mu \vec{K}}(\vec{r})=\frac{1}{\sqrt{N}} \sum_{\vec{R}_{n}}{e^{i\vec{K}\cdot\vec{R}_{n}} \chi_{\mu}(\vec{r}-\vec{R}_{n})}
	\end{equation}
	where $\chi_{\mu}$ is some atom-centered basis function, $\vec{K}$ is a crystal momentum vector similar to Eq.(D7-D9), and $N$ is the total number of unit cells. Using linear combination of atomic orbitals (LCAO), the periodic HF orbitals can be expressed as:
	\begin{equation}
	    \phi_{p \vec{K}}(\vec{r}) = \sum_{\mu}{C_{\mu p}(\vec{K}) \chi_{\mu \vec{K}}(\vec{r})}
	\end{equation}
	The corresponding eigenvalue equation at each $\vec{K}$ is therefore given by:
	\begin{equation}
	    \label{hf-pbc}
	    F(\vec{K}) C(\vec{K}) = S(\vec{K}) C(\vec{K}) E(\vec{K}),
	\end{equation}
    where the elements of Fock matrix $F$ and overlap matrix $S$ are given by:
    \begin{equation}
        F_{\mu \nu}(\vec{K}) = T_{\mu \nu}(\vec{K}) + V^{pp}_{\mu \nu}(\vec{K}) + J_{\mu \nu}(\vec{K}) - K_{\mu \nu}(\vec{K})
    \end{equation}
    \begin{equation}
        T_{\mu \nu}(\vec{K}) = -\frac{1}{2} \int_{\Omega} {\chi^{*}_{\mu \vec{K}}(\vec{r}) \nabla^{2}_{\vec{r}} \chi_{\nu \vec{K}}(\vec{r}) d\vec{r}}
    \end{equation}
    \begin{equation}
        J_{\mu \nu}(\vec{K}) = \int{ \int_{\Omega}{ \chi_{\mu \vec{K}}(\vec{r}) \frac{ \rho(\vec{r}', \vec{r}') }{ \abs{\vec{r} - \vec{r}'} } \chi^{*}_{\nu \vec{K}}(\vec{r}) d\vec{r}} d\vec{r}'}
    \end{equation}
    \begin{equation}
        K_{\mu \nu}(\vec{K}) = \int{ \int_{\Omega}{ \chi_{\mu \vec{K}}(\vec{r}) \frac{ \rho(\vec{r}, \vec{r}') }{ \abs{\vec{r} - \vec{r}'} } \chi^{*}_{\nu \vec{K}}(\vec{r}') d\vec{r}} d\vec{r}'}
    \end{equation}
    \begin{equation}
        S_{\mu \nu}(\vec{K}) = \int_{\Omega} {\chi^{*}_{\mu \vec{K}}(\vec{r}) \chi_{\nu \vec{K}}(\vec{r}) d\vec{r}}
    \end{equation}
    The $\Omega$ in the above equations represents integration in the unit cell. The corresponding one- and two-electrons in the Hamiltonian in Eq.(\ref{eq-ham}) can thus be computed from a Hartree-Fock calculation.

	\begin{acknowledgments}
		This work is supported by Central Research Institute, 2012 Labs, Huawei Technologies.
	\end{acknowledgments}

	\bibliography{citations.bib}
	
\end{document}